# The Fourth Partial Derivative In Transport Dynamics

Trinh, Khanh Tuoc

Institute of Food Nutrition and Human Health

Massey University, New Zealand

*K.T.Trinh@massey.ac.nz*

**Abstract**

A new fourth partial derivative is introduced for the study of transport dynamics. It is a Lagrangian partial derivative following the path of diffusion, not the path of convection. Use of this derivative decouples the effect of diffusion and convection and simplifies the analysis of transport processes.

## Introduction

The study of transport dynamics is based on partial differential equations that describe the variations of heat, mass and momentum fluxes with respect to time and space. The solution of these equations is particularly difficult because of their non-linearity and no solution yet exists for turbulent transport. Exact solutions exist only for laminar flow, conduction and diffusion especially in simple geometries. Eulerian solutions are based on fixed coordinates and can be used to analyse process operations and equipment. Lagrangian solutions are in principle simpler to derive because they are based on coordinates that lock onto the moving element. Thus the convective terms, the main source of the non-linear difficulties need not be considered in the analysis. However they introduce new problems: defining and tagging the moving element whose properties are modified by the transport processes as it moves along the convection path and relating the Lagrangian coordinates to the Eulerian coordinates. Most of the data required to verify theoretical derivations are obtained with fixed probes although recent advances in particle imaging velocimetry PIV hold the promise of true Lagrangian data. However PIV is expensive both in terms of resources and time so that Lagrangian data will always be restricted both in scope and volume. In fact Lagrangian fluid dynamics are so difficult to apply in practice that the author has only found one recent book on the subject (Bennett, 2006).

This paper introduces to the open literature a new partial derivative.

## Basic considerations

Attempts to linearise the transport equations started soon after their inception. For example the Navier-Stokes equations can be written for the unsteady flow pat a flat plate with a zero pressure gradient as

$$\frac{\partial u}{\partial t} + u\frac{\partial u}{\partial x} + v\frac{\partial u}{\partial y} = \nu\frac{\partial^2 u}{\partial y^2} \qquad (1)$$

Where $\nu$ is the kinematic fluid viscosity. For flow impulsively started Stokes (1851) neglected the convection terms to obtain

$$\frac{\partial u}{\partial t} = \nu\frac{\partial^2 u}{\partial y^2} \qquad (2)$$

Stokes has solved this equation for the conditions:

| | | | |
|---|---|---|---|
| IC | t = 0 | all y | $u = U_\infty$ |
| BC1 | $t > 0$ | y = 0 | u = 0 |
| BC2 | $t > 0$ | y = ∞ | $u = U_\infty$ |

where $U_\infty$ is the approach velocity for this sub-boundary layer. The velocity at any time t after the start of a period is given by:

$$\frac{u}{U_\infty} = erf(\eta_s) \qquad (3)$$

where $\eta_s = \frac{y}{\sqrt{4\nu t}}$

Many authors (Black, 1969; Einstein & Li, 1956; Hanratty, 1956; Meek & Baer, 1970) have successfully correlated the velocity profile in the wall layer of turbulent pipe flow with the Stokes solution.

Similarly the governing equation for heat transfer can be simplified for a laminar boundary layer past a flat plate to

$$\frac{\partial \theta}{\partial t} + u\frac{\partial \theta}{\partial x} + v\frac{\partial \theta}{\partial y} = \alpha\frac{\partial^2 \theta}{\partial y^2} \qquad (4)$$

where $\theta$ is the temperature, $\rho$ is the density and $\alpha$ the thermal diffusivity of the fluid. Higbie (1935) neglected the convection terms to develop a so-called penetration theory of transport

$$\frac{\partial \theta}{\partial t} = \alpha\frac{\partial^2 \theta}{\partial y^2} \qquad (5)$$

Assuming a third order polynomial temperature profile

$$\frac{\theta - \theta_w}{\theta_m - \theta_w} = 1.5\eta_h - \frac{1}{2}\eta_h^3 \qquad (6)$$

where

$$\eta_h = \frac{y}{\delta_h(t)} \qquad (7)$$

The thermal boundary layer thickness at time t is given by

$$\delta_h(t) = 2(2\alpha t)^{1/2} \qquad (8)$$

The rate of heat transfer is

$$q_w = 0.53 k \Delta\theta_\infty (\alpha t)^{1/2} \qquad (9)$$

Higbie estimated the time scale as

$$t = \frac{x}{U_\infty} \qquad (10)$$

Equation (9) successfully correlated the rate of absorption of a gas into a still liquid during short periods of exposure. However Trinh and Keey (1992) pointed out that equation (5) describes essentially the diffusion of a thermal front into the main flow and that a suitable time scale must be based on the thickness of the thermal boundary layer, not the longitudinal velocity and distance. Thus the focus of the analysis is basically turned by a 90° angle. When this is done, the penetration theory of Higbie is shown to transform exactly into the classic solution for heat transfer from a flat plate to a laminar boundary layer derived by Polhausen (1921). Thus in mathematical physics agreement between the theoretical derivation is not sufficient, the visualisation with its associated assumptions must be compatible with the physical reality.

The first concern lay with the Eulerian framework of the Stokes solution and its application to the wall layer comes from the classic observations of (Kline, Reynolds, Schraub, & Runstadler (1967), Despite the prevalence of viscous diffusion of momentum close to the wall, the flow was not laminar in the steady state sense envisaged by Prandtl (1935). Instead the region near the wall was the most active in the entire flow field. In plan view, Kline et al. observed a typical pattern of alternate low – and high-speed streaks. The low-speed streaks tended to lift, oscillate and eventually eject away from the wall in a violent burst. In side view, they recorded periodic inrushes of fast fluid from the outer region towards the wall. This fluid was then deflected into a vortical sweep along the wall. The low-speed streaks appeared to be made up of fluid underneath the travelling vortex as shown in Figure 1. The bursts can be compared to jets of fluids that penetrate into the main flow, and get slowly deflected until they become eventually aligned with the direction of the main flow. These observations have been confirmed by many others (Corino & Brodkey, 1969; Kim, Kline, & Reynolds, 1971; Offen & Kline, 1974).

The low speed streaks occur at seemingly random points in time and space. Yet the Stokes solution used by Einstein and Li and others is based on an Eulerian governing equation (2) . Is that reasonable? Mathematically speaking, a Lagrangian solution of the NS equation, even a subset, seemed to offer attractive advantages.

The second concern lay with the way the Stokes solution was derived. Stokes neglected the convection terms to obtain equation (1) which was to apply to a flat plate suddenly set in motion. Applying this solution to the sweep phase of the wall layer is not without controversy since the time scale is about 700 microseconds according to the data of Meek and Baer (1970) and the convection terms not quite negligible. Similarly since the Higbie solution can be transformed into the Polhausen solution for heat transfer, we need to justify the neglect of the convection terms in equation (5). Unfortunately neither Stokes, Einstein Li or Higbie provided proof that the convection terms are small enough to be neglected.

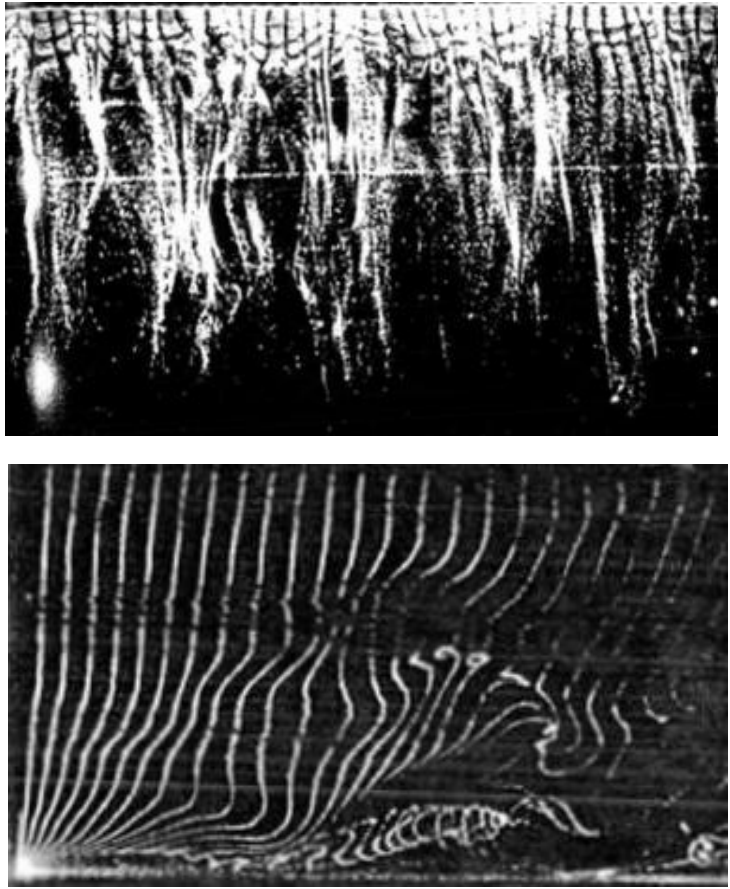

Figure 1. Hydrogen bubble visualisation of the wall process in turbulent flow. Top: low speed streaks after Kline et al. (1967), bottom side view of developing sub-boundary layer and burst after Kim et al. (1971).

Equation (4) may be rewritten as

$$\frac{D\theta}{Dt} = \alpha \frac{D^2\theta}{Dy^2} \qquad (11)$$

where

$$\frac{D\theta}{Dt} = \frac{\partial \theta}{\partial t} + u\frac{\partial \theta}{\partial x} + v\frac{\partial \theta}{\partial y} \qquad (12)$$

is called the substantial derivative.

Bird, Stewart, & Lightfoot (1960) p.73 illustrate the difference between the Eulerian partial derivative ∂θ/∂t and the Lagrangian substantial derivative Dθ/Dt with the following example. Suppose you want to count the fish population in a river. The Eulerian partial derivative gives the rate of change in fish concentration at a fixed point (x,y) in the river as seen by an observer standing on the shore. An observer in a boat drifting with the current will see the change in fish concentration on the side of the boat as given by the substantial derivative.

However, a fish swimming in the river will have a different perception of the fish population, which not described by either of these derivatives. Clearly a Lagrangian derivative is required but the convection velocities u and v are no longer relevant to this case; the velocity and path of the fish are. The introduction of this new kind of derivative, which is clearly needed, lies at the centre of this discussion (K.T. Trinh, 2002).

Consider for example the transport of heat from a flat plate to a laminar boundary layer in Figure 2. When we think of it, the flow of heat can be described physically as an unsteady state phenomenon, even when the momentum and thermal boundary layers profiles for laminar flow are steady. For a flat plate, each element of fluid moves from the leading edge only once and will experience a change in momentum along a streamline due to the effect of diffusion of viscous momentum from the wall. Similarly each thermal front generated from the wall also moves across the boundary layer only once. For convenience we will refer to a small portion of this front as an "entity" of heat. Thus the motions of these elements of fluid and entities of heat can be described with unsteady state equations.

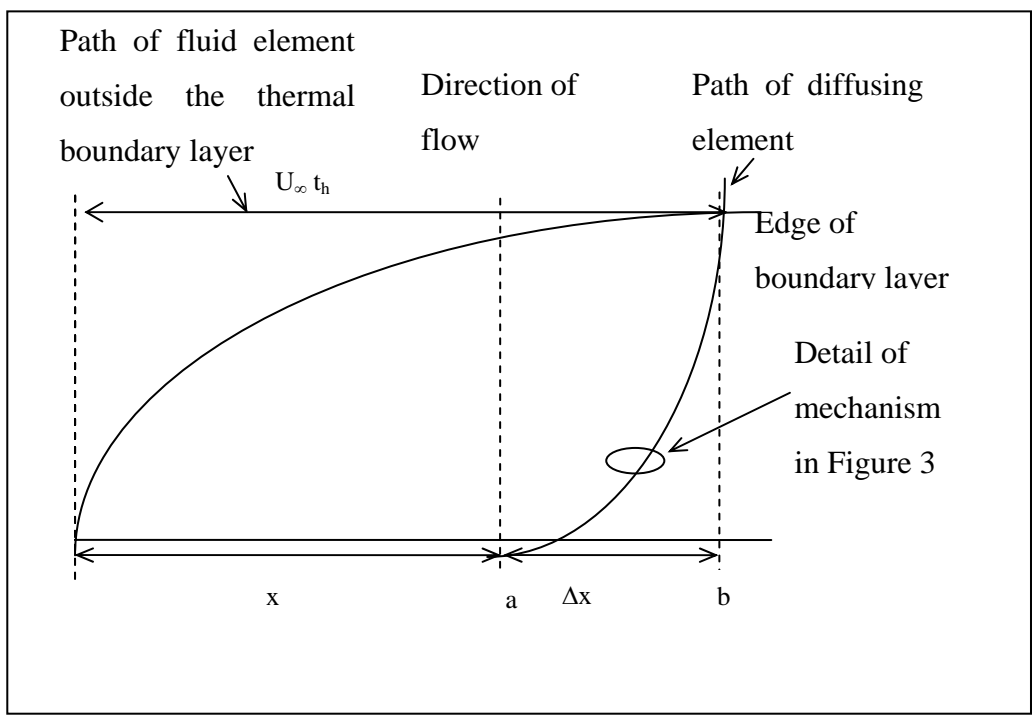

Figure 2  Diffusion path and convection velocities of an entity of heat after Trinh and Keey (1992)..

The appearance of steady state profile profiles is perceived in an Eulerian framework because of an endless repetition of unsteady state movements of elements that must be described in a Lagrangian framework. The movement of the elements of fluid and entities of heat are in different directions.

Consider next the nature of forces acting on an entity of heat. At time t, the entity of heat enters an element of fluid at position (x, y), drawn in full line and coloured red in Figure 3, which has velocities, u and v. At time (t + δt), the element of fluid has moved by convection to a new position, not shown in Figure 3, and the thermal entity has diffused to an adjacent element at (x + δx, y + δy), drawn in dotted lines and coloured orange with a brick pattern. *Since the thermal entity is a scalar property, it has no mass*, feels the effect of diffusion forces only and convects with the velocity of the fluid element where it temporarily resides. *The convective forces act on the host element of fluid, not on the entity of heat.*

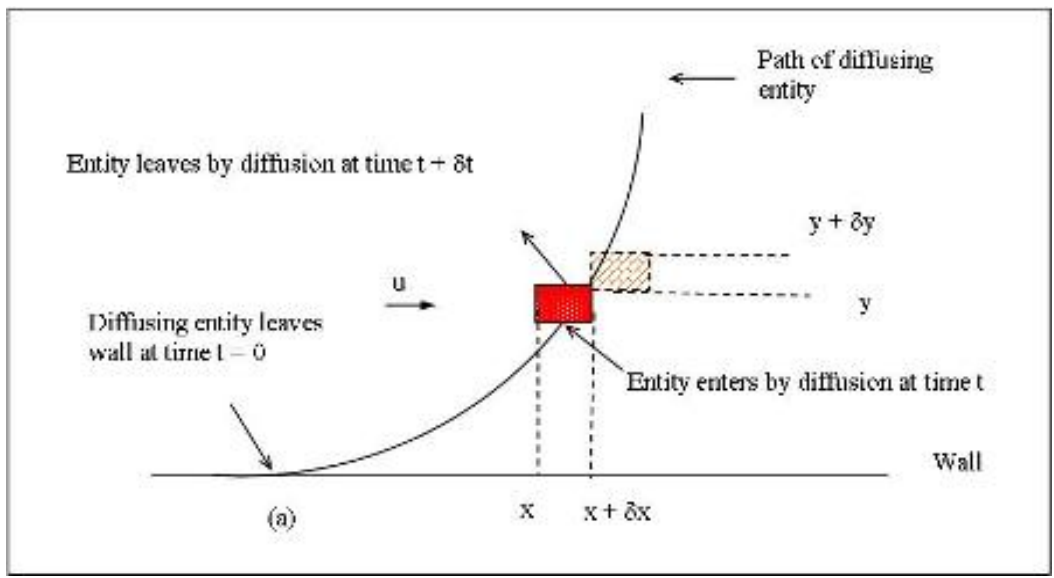

Figure 3. Convection and diffusion of an entity of heat

This physical analysis shows that the elements of fluid and heat move in different directions owing to different driving forces and it should be possible to separate the effects of diffusion and convection in the mathematical analysis. Consider now the mathematical description of this physical visualisation.

The fish in the Bird et al. example is represented here by the entity of heat illustrated in Figure 3. An observer attached to the *entity of heat* moving across the boundary layer will perceive the changes in temperature according to an equation similar to equation (5) but the frame of reference in the penetration theory is not attached to the wall as implied by Higbie (1935).

Partial derivatives were first formalised by Legendre and Jacobi in the nineteenth century (Wikipedia, 2009). Eulerian and Lagrangian derivatives were named after two contemporaries of Legendre and Jacobi who made important contributions to the field of mathematics but were use the same symbol. A new symbol $D/Dt$ is introduced for this new Lagrangian derivative *along the path of diffusion*. For a laminar boundary layer on a flat plate we obtain:

$$\frac{D\theta}{Dt} = \alpha \frac{D^2\theta}{Dy^2} \qquad (13)$$

This equation has the same form as the Fourier equation for heat transfer e.g. (Incropera, Dewitt, Bergman, & Lavine, 2007) that applies classically only to stationary media, but equation (13) can monitor the diffusion of heat into a convection stream along the path of thermal diffusion and give a more formal basis to penetration theories.

The decoupling of the flow paths of heat and fluid elements can also be found in the analysis of some well established unit operations. Consider for example the working of a multiple effect evaporator. For example in multiple effect evaporation latent heat enters with the fresh steam fed to the first effect, leaves it as it condenses and passes to the solution to be concentrated where it evaporates the solvent. This vapour then passes to the second effect as the new source of heat. Thus clearly the heat is reused many times but the condensed steam is drained after each effect (K. T. Trinh, 2009) p.116.

It must be clearly understood that this new partial derivative along the diffusion path of transport quantities is not the same as the traditional Lagrangian derivative (Bennett, 2006) that follow the path of convection. Such Lagrangian derivatives capture the changes to the fluid elements along their convection path.

## Validation

One way of validating the existence of this new partial derivative is of course to show that it can predict accurately well known transport problems. We consider here the problem illustrated in Figure 2 for a specific example detailed in Table 1but compare the predictions of the new Lagrangian solutions with well known Eulerian solutions that are widely accepted, rather than directly with experimental measurements.

Table 1. Parameters used in example of laminar flow and heat transfer past a flat plate

| Test fluid used | Water | unit |
|---|---|---|
| Temperature | 20 | °C |
| Viscosity | 0.001002 | kg/ms |
| Density | 998 | kg/m$^3$ |
| Heat capacity | 4.182 | kJ/kgK |
| Thermal conductivity | 0.603 | W/mK |
| Prandtl number | 6.935296 | |
| Approach velocity | 0.2 | m/s |

The physical properties were taken from Bayley, Owen, & Turner (1988). We follow an element of heat through the thermal boundary layer. The plate temperature from a position $x_0 = 0.05m$ to 25°C for a 5°C driving force. We monitor a heat element starting at the plate at the position $x = 0.10m$.

We divide the flow field into a rectangular grid. The velocity at each of the node points of that grid is calculated from the Polhausen approximate solution for laminar flow past a flat plate (Polhausen, 1921). The velocity associated with this rectangular element of fluid is simply taken in this example as the arithmetic average of the velocities at the four corners of the grid point

$$\bar{u} = \left( \frac{u(x,y) + u(x, y+\delta y) + u(x+\delta x, y+\delta y) + u(x+\delta x, y)}{4} \right) \quad (14)$$

We then calculate the time elapsed

$$\delta t = \frac{\delta x}{\bar{u}} \quad (15)$$

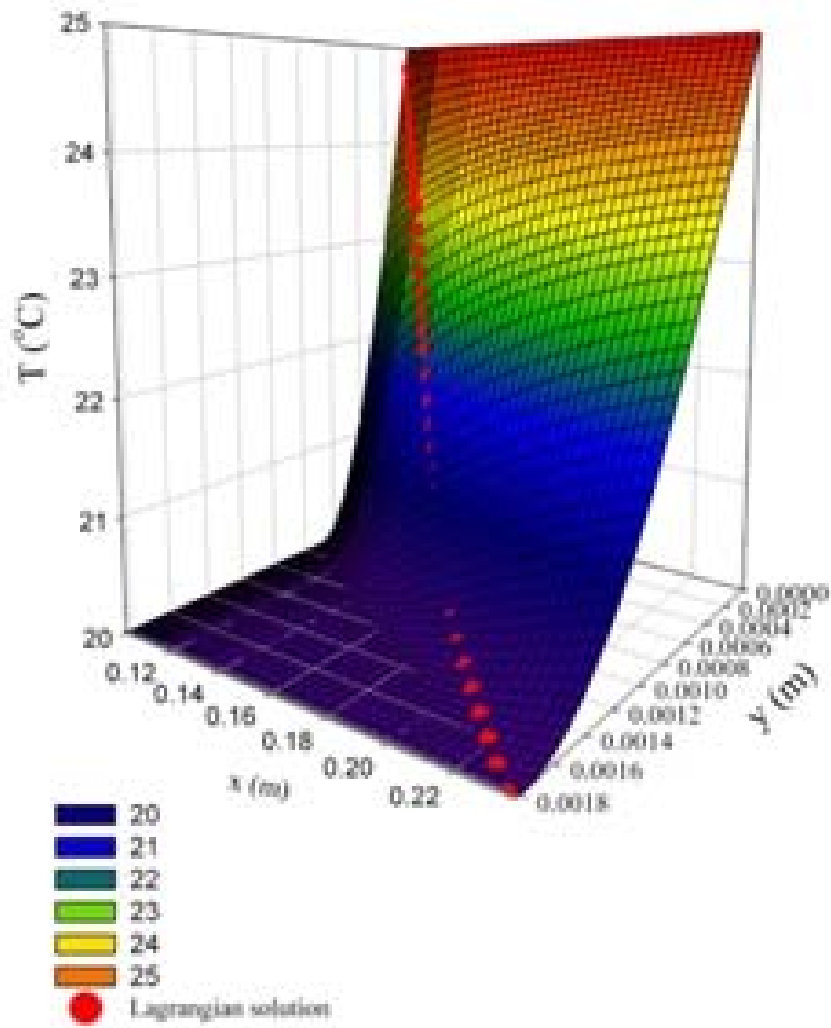

Figure 4 Diffusion path for a heat element starting at the wall at x=0.1 m and comparison with the Polhausen approximate solution for heat transfer in a laminar boundary layer

Thus for the first cell with $\delta x = 0.00125m, \delta y = 1.64337E-05$ the convection velocity of the fluid element is 0.003738359m/s, the time increment associated with diffusion of heat between points (0.10000,0) and (0.10125, 1.64377E-05) is 0.334371309s. The temperature at position (0.10125, 1.64377E-05) is then obtained from the solution of equation (6)

$$\frac{\theta_w - \theta}{\theta_w - \theta_\infty} = erf\left(\frac{y}{\sqrt{4\alpha t}}\right) \tag{16}$$

giving θ =24.78933904°C, The next element with corners (0.10125, 1.64377E-05), (0.1025, 1.64377E-05), (0.10125, 3.28754E-05) and (0.1025, 3.28754E-05) has velocity 0.011145875m/s with a time increment 0.112149s. Thus point (0.1025, 3.28754E-05) is associated with a time elapsed of (0.334371309 + 0.112149 =0.446520419s) giving a temperature of 24.63574535°C. This procedure allows us to follow the diffusion of heat between adjacent fluid elements.

The results of the Polhausen approximate solution for steady state (Eulerian) heat transfer in a laminar flat plate boundary layer (Polhausen, 1921; Squire, 1942) are presented in a 3D mesh plot in Figure 4. The results from equation (16) are overlaid as a 3D scatter plot. They fall almost exactly on the response surface of the mesh plot. In fact the standard deviation between the 38 points generated in the Lagrangian solution (scatter plot) and the corresponding sets of (x,y) in the Polhausen solution is 0.01%. The edge of the boundary layer in the mesh plot is clearly seen where the response surface flattens to a horizontal plane ($\theta = 20^\circ C$ coloured dark purple). The red scatter plot shows that the Lagrangian solution does indeed predict the boundary layer thickness accurately but supports and illustrate the argument of Trinh and Keey (op.cit.) that the thickness of thermal boundary which began at $x_o = 0.05m$ in this example, is not situated vertically above the position where the element of heat left the plate ($x_1 = 0.10m$) but at a distance much further downstream ($x = 0.235m$) in this case. Simultaneously, we obtain an estimate of the period of for the diffusion process ($t_d = 1.207464\,s$) and can calculate the rate of penetration of the thermal front, which is not available in the Eulerian solution. This extra information can be useful in many applications, for example in mixing processes.

## Discussion

The value of $D\theta/Dt$ at the point (0.125, 0.000328754) is given by our data base as

8.948451. The average value of Eulerian convective term $u(\partial\theta/\partial x)$ between points (0.1225, 0.000328754 and 0.127523, 0.000328754) is 4.852299. Clearly the convection term cannot be neglected in the Eulerian equation (4) to give equation (5) with physical realism.

While equations (2), (11) and (13) have similar forms, only $\partial\theta/\partial t$ and $D\theta/Dt$ are truly partial derivatives of the variable θ with respect to time. The substantial derivative Dθ/Dt is not. Thus similar techniques of analysis may be applied to equations (2) and (13) provided one remembers that the frames of reference for the solutions are different. The same is true for the traditional Lagrangian analysis.

This derivative allows us to extract from the NS equations subsets that are much more solvable because of the reduced problems of non linearity as the convection terms are omitted. For example, the wall layer can be analysed with the help of the Stokes solution

$$\frac{D\tilde{u}}{Dt} = \nu \frac{D^2\tilde{u}}{Dy^2} \qquad (17)$$

Where $\tilde{u}$ is the smoothed velocity of the sweep phase without the need for any empirical model. This in turns gives a method for a theoretical closure of the Reynolds-Averaged Navier-Stokes equations. These developments and further applications to turbulent transport are discussed in more detail in Trinh (2009)

While the derivative was originally developed to simplify the analysis of transport dynamics by decoupling the diffusion and convection processes, the same principle can of course be applied to other disciplines. Consider the dissemination of knowledge among different civilisations. There are two elements to this process: the transfer of information from one person to another (diffusion) and the transport of that information to another location (convection), We can also decouple the movement of the message (diffusion) and the medium (convection). In older days this knowledge transfer was convection controlled.. For example the use of coal in the west was introduced by Marco Polo. First Marco Polo had to learn the technology in China (diffusion of knowledge) then return to Italy (convection of the host element) before he could pass it to others in the West. With improvements in transport facilities from trains and ships to planes the control of this dissemination process has shifted from convection to diffusion not because the diffusion process (passing information from one individual to another) has improved significantly but because the convection term has changed dramatically. The medium can also change with scientific progress from people with oral traditions, to book with written words. With the invention of the internet, the medium has changed further to electric pulses and tipped the control further towards the diffusion step. We can thus create quantitative models of knowledge dissemination among different civilisations by modelling separately their handling of the convection and diffusion processes before recoupling the results.

## Conclusion

There exist a new Lagrangian partial derivative following the path of diffusion. Use of this derivative decouples the effect of diffusion and convection and simplifies the analysis of transport processes.